\documentclass[12pt]{article}
\usepackage{graphicx}
\usepackage[authoryear,longnamesfirst]{natbib}
\usepackage{amssymb}
\usepackage{lipsum}
\usepackage{url}
\usepackage{graphicx}
\usepackage{amsmath, esint}
\usepackage{color, hyperref, natbib}
\usepackage{lipsum}
\usepackage{epstopdf}
\usepackage{lineno}
\usepackage{caption}
\usepackage{fullpage}
\usepackage{authblk}

\def\tsc#1{\csdef{#1}{\textsc{\lowercase{#1}}\xspace}}
\tsc{WGM}
\tsc{QE}
\tsc{EP}
\tsc{PMS}
\tsc{BEC}
\tsc{DE}

\title{\textbf{Two-Crested Stokes Waves}}
\author{\Large A.~Semenova\thanks{Corresponding author: asemenov@uw.edu}}
\affil{\textit{\small Department of Applied Mathematics, University of Washington, Seattle, WA 98195-3925}}
\date{}

\begin{document}


\maketitle

\begin{abstract}
\noindent We study two-crested traveling Stokes waves on the surface of an ideal fluid with infinite depth. Following Chen~\&~Saffman (1980), we refer to these waves as class $\mathrm{II}$ Stokes waves. The class $\mathrm{II}$ waves are found from bifurcations from the primary branch of Stokes waves away from the flat surface. These waves are strongly nonlinear, and are disconnected from small-amplitude solutions. Distinct class $\mathrm{II}$ bifurcations are found to occur in the first two oscillations of the velocity versus steepness diagram. 
The bifurcations in distinct oscillations are not connected via a continuous family of class $\mathrm{II}$ waves. We follow the first two families of class $\mathrm{II}$ waves, which we refer to as the secondary branch (that is primary class $\mathrm{II}$ branch), and the tertiary branch (that is secondary class $\mathrm{II}$ branch). Similar to Stokes waves, the class $\mathrm{II}$ waves follow through a sequence of oscillations in velocity as their steepness rises, and indicate the existence of limiting class $\mathrm{II}$ Stokes waves characterized by a $120$ degree angle at every other wave crest.
\end{abstract}

\section*{Introduction}
Nonlinear traveling waves on the surface of a $2$D ideal fluid are called Stokes waves, first described by~\cite{stokes1847theory,stokes1880theory}. Such waves retain their shape in the reference frame moving with their velocity. Stokes waves, and in particular the ones with one crest per period have been studied extensively. Following~\cite{chen1980numerical} we refer to such waves as class $\mathrm{I}$ waves. Specifically, the existence of such waves was shown by~\cite{nekrasov1921waves} and~\cite{levi1925determination}. It was proven by~\cite{toland1978existence} that there exists a limiting class $\mathrm{I}$ Stokes wave, and for the proof that an angle of $120$ degrees forms at its crest, see~\cite{amick1982stokes, plotnikov1982}. 
The study of the stability of Stokes waves has a long history, see for example~\cite{benjamin1967disintegration}, \cite{longuet1997crest}, 
\cite{nguyenstrauss}, \cite{berti2022full}, \cite{Deconinck_Dyachenko_Semenova_2024}. In the work~\cite{deconinck2023instability}, it was shown that large, near limiting waves are unstable with respect to disturbances localized at their crests. In particular, it was found that the dominant disturbances are either co-periodic or have twice the period of the wave. 
The subject of this paper is traveling waves that have two crests per wavelength referred to as the class $\mathrm{II}$ Stokes waves.

We consider a $2$D ideal fluid with a free surface and infinite depth.   
The flow is assumed to be potential with fluid velocity $\textbf{v}=\nabla \phi$, where $\phi$ is the velocity potential.
The only force acting on the fluid is gravity, and periodic traveling surface waves  
are considered. 
The free surface is described by a $1$D curve $y = \eta(x,t)$, where $x$ and $y$ are horizontal and vertical spatial variables, and $t$ is time.
The motion of the free surface is described by the Laplace  
equation in the fluid domain $\mathcal{D} = \{(x,y)| -\pi< x < \pi, -\infty < y < \eta(x,t)\}$ and nonlinear time-dependent boundary conditions,
\begin{align*}
    &\begin{cases}
    \Delta \varphi = 0 \text{ in } \mathcal{D}, \\
    \eta_t = -\varphi_x \eta_x+\varphi_y \text{ at } y = \eta (x,t), \\
    \varphi_t +\frac{1}{2} \left( \nabla \varphi \right )^2 +g\eta = 0 \text{ at } y = \eta (x,t), \\
    \end{cases}
    \left.\varphi_y\right|_{y \to -\infty} = 0.
\end{align*}
Here $g$ is the acceleration due to gravity. 
\begin{figure}[ht!]
  \centering
  \includegraphics[width=0.9\textwidth]{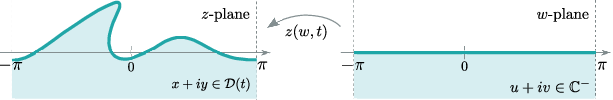}
  \caption{The region in the $w$ plane $(u,v) \in [-\pi,\pi]\times(-\infty,0]$ is mapped into the domain occupied by  the fluid in the $(x,y)-$plane $(x,y)\in [\pi, \pi]\times(-\infty, \eta(x,t)]$.   The line $v=0$ (blue line) is mapped onto the free surface $\eta(x,t)$.}
  \label{confmap}
\end{figure}
We use the conformal transformation approach described in~\cite{ovsyannikov1973dynamika} and 
later in~\cite{tanveer1991singularities,dyachenko1996dynamics} 
(see also~\cite{dyachenko2016branch} for the periodic problem). 
In the conformal variables the equations of motion can be found from the variational principle, and have the form,
\begin{align}
&y_t x_u - y_u x_t = -\hat H\psi_u, \label{imp1}  \\
&x_t \psi_u - x_u \psi_t - \hat H\left[y_t \psi_u -  y_u\psi_t \right] =
g\left(x_uy - \frac{1}{2} \hat H \partial_u y^2\right), \label{imp2}
\end{align}
where $\hat H$ is the Hilbert transform described below, and $\psi$ is the value of the potential $\varphi$ on the surface. The conformal map $z(w,t) = x(w,t)+ i y(w,t)$ is a time-dependent holomorphic function of $w=u+iv$ that maps the lower complex half-plane $(u,v) \in (-\pi,\pi)\times(-\infty,0]$ into the fluid domain $\mathcal{D}$ as presented in Figure~\ref{confmap}. The equations~\eqref{imp1}--\eqref{imp2} are posed on the real line $v = 0$, and $\hat H$ is the periodic Hilbert transform that is defined by,
\begin{align}
	\hat Hf(u)
= \frac{1}{2\pi} \fint^{\pi}_{-\pi} f(u') \cot{\frac{u'-u}{2}}du'.
\end{align}
In Fourier space, the Hilbert transform has the Fourier symbol $i\,\mbox{sign}{k}$, or equivalently  $ \hat H e^{iku} = i\,\mbox{sign}(k)\, e^{iku}$. The Hamiltonian of the fluid is expressed in the conformal variables as,
\begin{align}
    \mathcal{H} = \frac{1}{2}\int_{-\pi}^{\pi} \psi \hat k \psi \,du + \frac{g}{2}\int_{-\pi}^{\pi} y^2 x_u \,du,\label{HamConf}
\end{align}
where $\hat k$ is the Dirichlet-to-Neumann operator with $\hat k = -\hat H \partial_u$ for the lower complex half-plane.

We substitute a traveling wave ansatz $y=y(u-ct)$, $\psi=\psi(u-ct)$ with propagation velocity $c$ into the equations~\eqref{imp1}--\eqref{imp2}, which yields the
~\cite{babenko1987some} equation,
\begin{align}
	\left(c^2 \hat k - g\right) y\left(u\right) - \frac{g}{2}\left[\hat ky^2\left(u\right) + 2y\left(u\right)\hat k y\left(u\right)\right] = 0.\label{Babenko}
\end{align}
Here the conformal map $z$ and the potential $\psi$ are uniquely determined from $y(u)$.

Waves with two distinct crests per wavelength have been computed by~\cite{chen1980numerical} and are called class $\mathrm{II}$ waves (here $\mathrm{II}$ stands for the number of crests per period). In that paper, the authors also studied three-crested waves which are referred to as class $\mathrm{III}$ solutions. 
The bifurcation points from the family of one-crested Stokes waves to symmetric multi-period waves were also studied by~\cite{saffman1980long, vanden1983some,longuet1985bifurcation, vanden2017new}.
In this text, we adopt the~\cite{chen1980numerical} terminology and refer to the solutions of equation~\eqref{Babenko} with two crests of different amplitudes per period as class $\mathrm{II}$ waves. We also introduce notation for a Stokes wave with one crest per period as the solution from the main branch, or a class $\mathrm{I}$ wave.
Non-symmetric waves and corresponding bifurcation points on the main branch of Stokes waves are discussed by~\cite{zufiria1987non}, and recently by~\cite{wilkening2021spatially}.

Similarly to~\cite{DyachenkoLushnikovKorotkevichJETPLett2014}, we solve the equation~\eqref{Babenko} via the Newton-MINRES method (see~\cite{yang2009newton}) in Fourier space. We have chosen to use the minimal residual method (MINRES) instead of the Conjugate-Gradient method because although the linearized Babenko operator is self-adjoint, it is not positive definite.
We use the double period bifurcation points 
computed in~\cite{dyachenko2023quasiperiodic} for $2\pi$-periodic Stokes waves to find and compute traveling waves that have twice the period of the original one-crested Stokes waves. 
Since we consider class $\mathrm{II}$ waves with wavelength $2\pi$, we rescale a ``one crested Stokes wave'' 
with wavelength $2\pi$ to have two identical crests of half amplitude in each interval of length $2\pi$. 
Given a $2\pi$-periodic one-crested Stokes wave with velocity $c$, and a vertical displacement $y(u)$, the wave of the form $\left(c/\sqrt{2}, y(2u)/2 \right)$ is also a solution of the Babenko equation. It has exactly two identical crests on the interval $u \in \left[-\pi,\pi\right]$. 
When a Stokes wave is located at a double period bifurcation point, we can compute and study properties of $2\pi$-periodic class $\mathrm{II}$ Stokes waves by means of the continuation method in the velocity parameter, $c$, exactly as we would continue from the flat water to the main branch of class $I$ Stokes waves.

To compute  class $\mathrm{II}$ waves, we use the continuation method for $c$ and start from the velocity $c^s$. The initial guess for the Newton-MINRES iterations is chosen to be 
\begin{equation}
y_0(u) = y^s(u)+\varepsilon f(u), \label{pert}
\end{equation}
where $\left(c^s, y^s(u)\right)$ is the solution of the Babenko equation~\eqref{Babenko} from the primary branch at the period-doubling bifurcation point. 
Here $f(u)$ is the eigenfunction corresponding to the zero eigenvalue of the linearized Babenko operator (see~\cite{dyachenko2023quasiperiodic} for details), and $\varepsilon$ is a small number (we set $\varepsilon$ to be $10^{-2}$). 
With this choice of the initial guess, the Newton-MINRES method converges to a class $\mathrm{II}$ traveling wave of the Babenko equation~\eqref{Babenko}. 
The sign of $\varepsilon$ dictates which crest (left or right) becomes the taller one past the bifurcation point, however the resulting solution branches are equivalent class $\mathrm{II}$ waves and are identical after a horizontal shift by one half of a wavelength. Hence, without loss of generality the primary branch solution is perturbed via~\eqref{pert} with $\varepsilon>0$. 
The auxiliary conformal transformation from ~\cite{lushnikov2017new} is used for computations of steep waves to speed up convergence of the Fourier series. We observe that a corner of $120$ degrees tends to form at the taller crest as we follow along the branch, see Fig.~\ref{sol}.

\vspace{-0.1cm}
\section*{Waves Profiles and Discussion}
\begin{figure}
  \centering
  \includegraphics[width=0.99\textwidth]{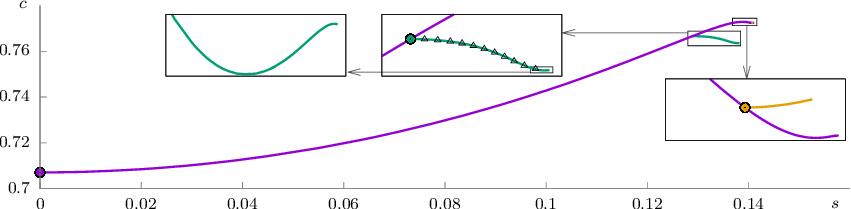}
  \caption{Oscillations in velocity $c$ as a function of steepness $s$. The purple curve corresponds to Stokes waves from the primary branch (one crested Stokes waves per wavelength); The green curve represents the class $\mathrm{II}$ Stokes waves with period $2\pi$ bifurcating from $s_1$ (secondary branch); The gold curve is the second family of the class $\mathrm{II}$ waves (tertiary branch) bifurcation from $s_2$. Black triangles are the data for class II waves from~\cite{chen1980numerical} which fits well onto the secondary branch (green curve).}
  \label{vel}
\end{figure}

~\cite{chen1980numerical} computed the first double period-bifurcation point $s_1^{CS} = 0.1289$ on the primary branch of the $2\pi$-periodic class $\mathrm{I}$ Stokes waves. 
The second bifurcation point $s_2^{VB} = 0.14$ was computed by~\cite{vanden1983some}. Recently,~\cite{vanden2017new} computed the third bifurcation point together with some class $\mathrm{II}$ waves from the quaternary branch. In this work, we use double-period bifurcation points $s_1 = 0.128903$, $s_2 = 0.140487$, and $s_3 = 0.141032049$ computed in~\cite{dyachenko2023quasiperiodic} and the conformal mapping approach described above to study the class $\mathrm{II}$ waves from the secondary and tertiary branches. 
We define the steepness parameter to be $s = H/\pi$ which is the ratio of the distance between the taller crest and the deeper trough $H$ divided by $\pi$. 
This parameter differs by a multiple of $1/2$ from the conventional definition of steepness which is the ratio of the height to the wavelength (in this case it is $2\pi$). 
The reason for this choice is the ability to compare the values of $s$ to the steepness of extensively studied one-crested $2\pi$-periodic Stokes waves.

\begin{figure}
  \centering
  \includegraphics[width=0.99\textwidth]{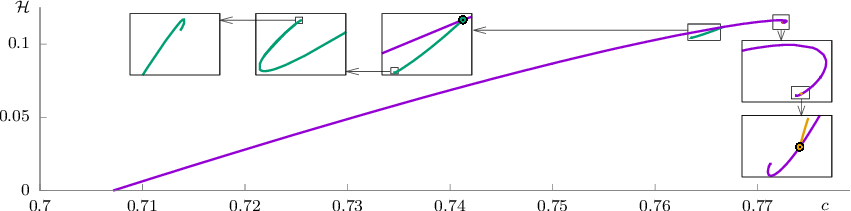}
  \caption{The Hamiltonian $\mathcal{H}$ as a function of velocity $c$. Green and gold circles mark the double-period bifurcation points from the primary branch of class $\mathrm{I}$ Stokes waves (purple curve). Class $\mathrm{II}$ waves from the secondary and tertiary branch are shown by green and gold curves respectively. The insets show winding of the primary (purple) and secondary (green) branches. These spirals indicate extrema in both the Hamiltonian and the velocity as functions of steepness for these $2$ curves.}
  \label{vel_H}
\end{figure}

\begin{figure}
  \centering
  \includegraphics[width=0.99\textwidth]{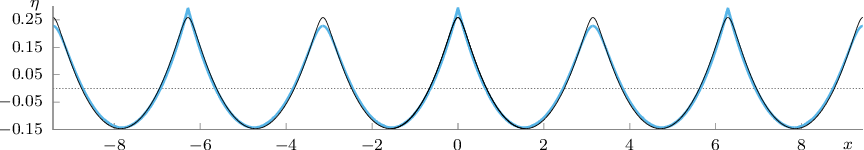}
  \includegraphics[width=0.99\textwidth]{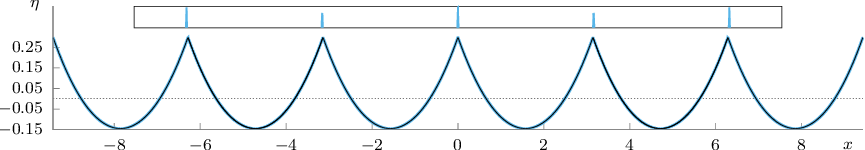}
  \caption{Both panels show $3$ periods of waves. The black lines are the Stokes waves from the primary branch, $s_1$ (top panel) and $s_2$ (bottom panel).
  (Top Panel) Profile of the class $\mathrm{II}$ wave $(s,c) = (0.1381671,0.7635778)$ (blue curve) from the secondary branch (originating from $s_1$).  
  (Bottom Panel) Profile of the class $\mathrm{II}$ wave $(s,c) = (0.1408973, 0.77245)$ (blue curve) from the tertiary branch (originating from $s_2$). 
  The difference in the crests amplitudes is shown in the inset.}
  \label{sol}
\end{figure}

In Figure~\ref{vel}, we show the velocity $c$ as a function of steepness $s$. The purple curve represents the class $\mathrm{I}$ Stokes waves from the primary branch with oscillations in $c$ as the limiting wave is approached, as we know from the works of~\cite{longuet1978theory, DyachenkoLushnikovKorotkevichJETPLett2014}. 
The green curve corresponds to the branch of $2\pi$-periodic waves of class $\mathrm{II}$ where one of the crests is taller than the other. We refer to this family of class $\mathrm{II}$ Stokes waves as the secondary branch. These waves bifurcate from the double-period bifurcation point $s_1 = 0.128903$ (marked by the green circle) computed  in~\cite{dyachenko2023quasiperiodic}. 
The black triangles are the data from~\cite{chen1980numerical} which fits well with our results for the secondary branch except for the last three waves. We are able to extend the results of~\cite{chen1980numerical} and~\cite{vanden2017new}, and compute even steeper waves from the secondary branch as can be seen in the two insets for the green curve that show oscillations in velocity as the steepness of waves increases. 
We compute $2$ local extrema (first minimum and second maximum) in the green curve, and conjecture that there are infinitely many such extrema, qualitatively similar to the class $\mathrm{I}$ waves on the main branch. Also, we note that a corner of $120$ degrees forms at the taller crest and a class $\mathrm{II}$ limiting wave is evident.
The limiting value of velocity $c$ for class $\mathrm{II}$ waves is $c^{2nd}_{lim} = 0.7635\ldots$ (with $4$ digits of accuracy) which is different from the value $c^p_{lim} = 0.77236216478\ldots$ for the limiting Stokes wave from the primary branch, see~\cite{lushnikov2017new}. 
The golden curve in Figure~\ref{vel} represents the second family of two-crested waves that form the tertiary branch and bifurcates from the primary branch of Stokes waves at $s_2 = 0.140487$. We conjecture that this branch also has infinitely many oscillations in the velocity as a function of the wave steepness and has its own limiting wave. The value of the velocity for this limiting wave will be different from the values for primary (purple curve) and secondary (green curve) branches. Moreover, we know that there is at least one more double-period bifurcation point $s_3 = 0.141032049$ computed in~\cite{dyachenko2023quasiperiodic}, and there is a third family of class $\mathrm{II}$ waves (the quaternary branch) originating from it, see~\cite{vanden2017new}. Furthermore, we conjecture that there are infinitely many such families/branches of class $\mathrm{II}$ waves (since it is conjectured in~\cite{dyachenko2023quasiperiodic} that there are infinitely many zero eigenvalues of the linearized Babenko operator). They all form a corner at one or more crests as their respective limiting waves are approached. 

The plot of the Hamiltonian $\mathcal{H}$ as a function of velocity $c$ is a spiral as we present in Figure~\ref{vel_H}.
The purple line represents Stokes waves from the primary branch, and it spirals around the center $\left(c_{lim}^p,\mathcal{H}_{lim}^p\right)$ corresponding to the limiting class $\mathrm{I}$ Stokes wave.
When shown as the function of wave steepness $s$ the Hamiltonian oscillates as the limiting wave is approached, see ~\cite{longuet1977theory, korotkevich2022superharmonic}, and the extrema of the velocity and the Hamiltonian correspond to the points where the tangent line to the spiral in Figure~\ref{vel_H} is vertical and horizontal respectively.
The extrema of the Hamiltonian occur before the extrema of the velocity and correspond to a clockwise spiral.
The green and gold circles on the primary branch mark the double-period bifurcation points to the secondary (green) and tertiary (gold) branches of class $\mathrm{II}$ waves respectively. 
The first two extrema in the Hamiltonian of the secondary branch are captured by the winding of the green curve and presented in the three insets. 
We conjecture that more bifurcation points exist on the secondary branch as the curve spirals towards the class $\mathrm{II}$ limiting wave $\left(c_{lim}^{2nd},\mathcal{H}_{lim}^{2nd}\right)$. 
Similar to the diagram for class $\mathrm{I}$ waves, we note that the local extrema in the Hamiltonian occur at a smaller steepness than the extrema in the velocity.
We expect the tertiary branch to have similar winding behavior to the primary and secondary branches, and spiral to its own limiting wave.

\newpage
\begin{figure}
    \centering
    \includegraphics[width=0.495\textwidth]{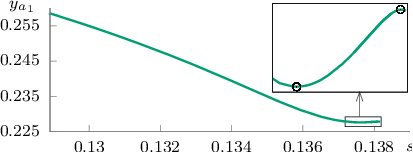}
    \includegraphics[width=0.495\textwidth]{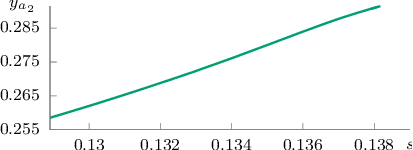}
    \caption{In both panels we plot the crests amplitudes of class $\mathrm{II}$ waves from the secondary branch as functions of the waves steepness. The amplitude of the smaller crest (Left Panel) oscillates (increases and decreases) and the amplitude of the crest that forms an angle of $120$ degrees (Right Panel) monotonically increases as the limiting class $\mathrm{II}$ wave is approached.
    The first $2$ extremizer of $y_{a_1}$ are $s^{*} = 0.137619$ and $s^{*} = 0.13814$ (marked by black circles).
    } 
    \label{crests}
\end{figure}

\begin{figure}
  \centering
  \includegraphics[width=0.99\textwidth]{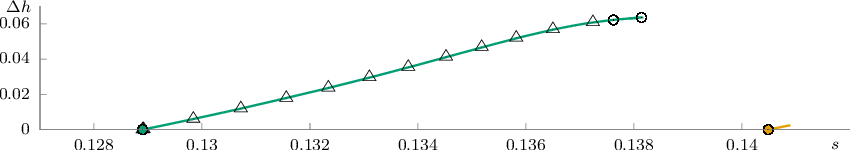}
  \caption{The height difference $\Delta h$ between two crests of class $\mathrm{II}$ waves versus steepness $s$ for the secondary branch (green curve) and tertiary branch (golden curve). Black triangles correspond to data from~\cite{chen1980numerical}. The green and golden circles mark the first and second double period bifurcation points ($s_1$ and $s_2$) respectively. 
  }
  \label{amp}
\end{figure}

In Figure~\ref{sol}, we present profiles of class $\mathrm{II}$ waves (blue curves) from the secondary (top panel) and tertiary (bottom panel) branches. 
The main branch solutions with steepness $s_1$ and $s_2$ correspond to Stokes waves that bifurcate to the class $\mathrm{II}$ waves, and are marked with black lines in the top and bottom panels of Figure~\ref{sol} respectively.
Three periods of class $\mathrm{II}$ waves with 
$s = 0.138167\ldots$ (top panel) and $s =  0.140897\ldots$ (bottom panel)  
are shown. 
The inset in the bottom panel is a zoom-in to the crest region of that wave, and is necessary to display the amplitude difference which is otherwise visually indistinguishable.
We track the amplitudes of the crests of class $\mathrm{II}$ waves from the secondary branch in Figure~\ref{crests}.
One of the crests grows monotonically and forms a corner (right panel) while the amplitude of the other one oscillates (left panel). We conjecture that there are infinitely many such oscillations in the amplitude of the other crest, and it does not form a corner as the steeper crest approaches the limiting shape.
The height difference between the two crests $\Delta h$ (green curve 
in Figure~\ref{amp}) grows as the steepness of the waves increases.
The black triangles correspond to the data from~\cite{chen1980numerical}, and they are marked to offer a direct comparison with preceding studies of double-period bifurcations (green and gold circles are the double period bifurcations).

\vspace{-0.1cm}
\section*{Conclusion}
We have computed and studied class $\mathrm{II}$ waves (traveling waves with two crest of different amplitudes per period) thus expanding observations of~\cite{chen1980numerical, vanden2017new}. Two branches of class $\mathrm{II}$ waves have been computed. The secondary branch bifurcates from $s_1 = 0.128903$, and it was shown that the velocity $c$ and the Hamiltonian $\mathcal{H}$ (for the secondary branch) have oscillations as the limiting class $\mathrm{II}$ wave $c^{2nd}_{lim} = 0.7635$ is approached.
The tertiary branch of two crested wave bifurcating from $s_2 = 0.140487$ is shown. We conjecture that the velocity $c$ and the Hamiltonian $\mathcal{H}$ oscillates for steeper waves from this branch just as it does for the class $\mathrm{I}$ waves, and the details of this behaviour is left for future work. 
Such recurrent behaviour, together with the conjecture in~\cite{dyachenko2023quasiperiodic} about an infinite number of zero eigenvalues of the linearized Babenko operator, may be viewed as a reinforcement that 
an infinite number of bifurcating points to class $\mathrm{II}$ waves is located at the primary branch of Stokes waves, see~\cite{vanden2017new}. The stability of the class $\mathrm{II}$ waves and comparison with results from~\cite{deconinck2023instability} for class $\mathrm{I}$ Stokes waves from the primary branch remains open and is left for future work.

\vspace{0.2cm}
\noindent {\bf Acknowledgements} The author thanks B. Deconinck and S. A. Dyachenko for helpful discussions. The author thanks the Pacific Institute for the Mathematical Sciences and Simons Foundation for support.

\bibliographystyle{plainnat}
\bibliography{water-waves-refs}

\end{document}